\documentclass{emulateapj}

\shorttitle{Near-Infrared Analysis of the Submillimeter Background}
\shortauthors{Wang, Cowie, \& Barger}

\begin{document}

\title{A Near-Infrared Analysis of the 
Submillimeter Background and the Cosmic Star-Formation History}

\author{W.-H. Wang,\altaffilmark{1}
L. L. Cowie,\altaffilmark{1}
and A. J. Barger\altaffilmark{2,3,1}}
\altaffiltext{1}{Institute for Astronomy, University of Hawaii, 2680 Woodlawn Drive, Honolulu, HI 96822}
\altaffiltext{2}{Department of Astronomy, University of Wisconsin-Madison, 475 North Charter Street, Madison, WI 53706}
\altaffiltext{3}{Department of Physics and Astronomy, University of Hawaii, 2505 Correa Road, Honolulu, HI 96822}

\slugcomment{Accepted by The Astrophysical Journal for v647n 1 issue, August 2006.}

\begin{abstract}
We use new deep near-infrared (NIR) and mid-infrared (MIR) observations to analyze 
the 850$~\mu$m image of the Great Observatories Origins Deep Survey-North
region around the Hubble Deep Field-North. We show that much of the 
submillimeter background at this wavelength is picked out by sources with 
$H(AB)$ or 3.6~$\mu {\rm m} (AB)<23.25$ (1.8 $\mu$Jy). These sources contribute an 
$850~\mu$m background 
of $24\pm2$~Jy~deg$^{-2}$. This is a much higher fraction of the measured background 
($31-45$~Jy~deg$^{-2}$) than is found with current 20~cm or $24~\mu$m samples. 
Roughly one-half of these NIR-selected sources have spectroscopic identifications, and we 
can assign robust photometric redshifts to nearly all of the remaining sources using their 
UV to MIR spectral energy distributions. 
We use the redshift and spectral type information to show that
a large fraction of the $850~\mu$m background light comes from sources 
with $z=0-1.5$ and that the sources responsible have intermediate 
spectral types. Neither the elliptical galaxies, which have no star formation, nor the 
bluest galaxies, which have little dust, 
contribute a significant amount of 850~$\mu$m light, despite the
fact that together they comprise approximately half of the galaxies in the sample. 
The galaxies with intermediate spectral types have a mean flux of 
$0.40\pm0.03$~mJy at $850~\mu$m and $9.1\pm0.3~\mu$Jy at 20~cm.

The redshift distribution of the NIR-selected 850~$\mu$m light lies well below that 
of the much smaller amount of light traced by the more luminous, radio-selected submillimeter 
sources. We therefore require a revised star-formation history with a lower star-formation 
rate at high redshifts. We use a stacking analysis of the 20~cm light in the NIR sample to 
show that the star-formation history of the total 850~$\mu$m sample 
is relatively flat down to $z\sim 1$ and that half of
the total star formation occurs at redshifts $z<1.4$.

\end{abstract}


\section{Introduction}
\label{intro}

The integrated extragalactic background light (EBL) is a measure of the history of 
the luminous energy production of the universe from both star formation and active 
galactic nuclei (AGNs). Directly emitted light is seen in the UV and optical, whereas
dust reradiated energy appears in the far-infrared (FIR) and submillimeter.
\emph{COBE} obtained detailed measurements of the EBL at FIR and submillimeter
wavelengths \citep[e.g.,][]{puget96,fixsen98}, showing that the total radiated emission 
reprocessed by dust in the FIR/submillimeter is comparable to the total measured optical 
EBL. However, to proceed further, we also need to know the redshift distribution of the
sources contributing to the submillimeter background, 
and this information has been extremely difficult to obtain.

In the last decade, the submillimeter/millimeter EBL has been resolved into discrete 
sources by deep surveys with the Submillimeter Common-User Bolometer Array (SCUBA) 
on the 15~m James Clerk Maxwell Telescope (JCMT) and with the Max-Plank Millimeter Bolometer 
array on the 30~m IRAM telescope.
Blank-field surveys have resolved sources in the 
$2-20$~mJy range that account for $\sim20-30$\% of the 850~$\mu$m EBL 
\citep[e.g.,][]{barger98,hughes98,barger99a,eales99,eales00,eales03,bertoldi00,
scott02,webb03,borys03,wang04}. With the help of strong lensing, surveys in
cluster fields have resolved sources over the $0.3-2$~mJy range that account 
for a further $45-65$\% of the 850~$\mu$m EBL \citep{smail97,chapman02,cowie02,knudsen05}. 
Together these surveys provide a nearly complete resolution of the background at 
850~$\mu$m. The ``typical'' source contributing to the 850~$\mu$m EBL
has a mean flux of about 
0.9~mJy and a median flux of about 0.6~mJy \citep{cowie02}. 

However, the redshift follow-up of the submillimeter sources has been very slow. Because 
of the large beam size ($15\arcsec$) of SCUBA and the optically-faint nature of the dusty
sources, identifying the optical and near-infrared (NIR) counterparts to the submillimeter 
sources is time consuming \citep[e.g.,][]{barger99b,ivison00}. To date, the most successful 
identifications of the submillimeter sources rely on the empirical correlation between the 
nonthermal radio emission and the thermal dust emission \citep[e.g.,][]{condon92}. 
Once the radio 
counterparts to the submillimeter sources are detected by radio interferometers, the 
redshifts of the sources can be crudely estimated using the radio-to-submillimeter 
flux ratios \citep{carilli99,barger00,hughes02,ivison02,chapman03b} or accurately measured
with optical spectroscopy \citep{chapman03a,chapman05}. The radio-identified sources are 
mostly bright ($\gg 2~$mJy) submillimeter sources at $z=1.5-3.5$,  
with properties similar to the local ultraluminous
infrared galaxies (ULIRGs; $L_{\rm IR}>10^{12}~L_{\sun}$, where $L_{\rm IR}$ is
the $8-1000~\mu$m infrared luminosity; see, e.g., \citealp{sanders96}).  
We note, however, that because of the $K$-correction and the sensitivity limit in the radio, 
only $\sim60\%$ of the bright submillimeter sources are identified in the radio \citep{barger00}. 
It is not known whether the remaining 40\% are at higher redshifts that simply
cannot be reached by current radio telescopes. 

Importantly, however, the properties and redshift 
distribution of the faint submillimeter sources that dominate the submillimeter EBL 
remain essentially unknown. The absence of any redshift information for more than 90\% 
of the 850~$\mu$m EBL represents a formidable uncertainty in determining the star-formation 
history, and this is what we aim to resolve in the present paper.

Like the radio emission, the mid-infrared (MIR) emission at $\gtrsim5~\mu$m could serve as 
another proxy to the submillimeter emission, since it also comes from dust. The MIR window 
has been opened by the Infrared Array Camera (IRAC, \citealp{fazio04}) and the Multiband 
Imaging Photometer for \emph{Spitzer} (MIPS, \citealp{rieke04}) on the 
\emph{Spitzer Space Telescope} \citep[e.g.,][]{huang04,serjeant04,ivison04,egami04}.  
MIPS should be sensitive to $z\lesssim1$ galaxies with infrared 
luminosities similar to local normal galaxies ($L_{\rm IR}\sim 10^{10}~L_{\sun}$, 
corresponding to $\sim0.1$~mJy at 850~$\mu$m) and to $z\lesssim3.5$ ULIRGs 
(i.e., typical of the 
bright submillimeter sources). Thus, MIPS should be able to detect the radio-identified 
submillimeter sources at $z\lesssim3.5$ and to provide a large sample of faint 
sources that are beyond the confusion limit of current submillimeter telescopes 
\citep[e.g.,][]{chary04}. However, as we shall show in this paper, even the 
extraordinarily deep MIPS data of the Great Observatories Origins Deep 
Survey-North (GOODS-N) \emph{Spitzer} Legacy Science Program in the Hubble Deep 
Field-North (HDF-N) region does not substantially identify the 850~$\mu$m EBL.

Remarkably, however, the combination of a $J$ or $H$-band sample (selected from
images obtained with the new generation of ground-based, wide-field NIR cameras) 
and the IRAC 3.6~$\mu$m sample does identify much of the 850~$\mu$m EBL. We show 
this using the $H$-band image of the GOODS-N region obtained by \citet{trouille06}. 
This result makes sense if the bulk of the sources contributing to the 850~$\mu$m EBL
are actually at lower redshifts and luminosities than those identified at the brighter 
submillimeter fluxes. Such sources have strong rest-frame optical/NIR counterparts
that are picked up in the NIR sample. We utilize the spectroscopic and photometric 
redshift information on our NIR sample to confirm this result. We find that more than 
half of the 850~$\mu$m EBL arises in sources with $z<1.5$ and that the sources that
are responsible have intermediate spectral types.
Neither the elliptical galaxies, which have no star formation, nor the bluest galaxies, 
which have little dust, contribute substantially to the 850~$\mu$m EBL, despite the fact 
that together they comprise approximately half of the sample.

This result has profound implications for our understanding of the star-formation 
history, lowering previous estimates of the high-redshift star formation rate
densities by factors of at least two. We analyze the star-formation history 
of our NIR sample using a 20~cm stacking analysis and compare this with the 
maximum star formation rate density at higher redshifts obtained directly from the 
submillimeter light. Together these show that the total star formation rate density 
peaks at a redshift at or just below one and is roughly flat at higher redshifts.

The paper is organized as follows. The submillimeter, NIR, MIR, optical, 
radio, and X-ray data are described in \S\ref{secsample}. The spectroscopic 
and photometric redshifts are discussed in \S\ref{secz}. 
The use of the NIR, MIR, and radio populations to identify the 
submillimeter background is discussed in \S\ref{secebl}, and the 
850~$\mu$m EBL identified by the NIR-sample is broken down by galaxy flux,
color, spectral type, and redshift.
The star-formation history is described in \S\ref{section_evolution}.
Our main results are summarized in \S\ref{secsummary}. Throughout the 
paper, we assume the WMAP cosmology: $H_0=71$~km~s$^{-1}$~Mpc$^{-1}$, 
$\Omega_M=0.73$, and $\Omega_{\Lambda}=0.27$ \citep{bennett03}.

\section{The Data Samples}
\label{secsample}

\subsection{Submillimeter Data}
\label{secsmm}

The GOODS-N submillimeter map of \citet{wang04}
is based on jiggle-map data taken primarily by our group using the SCUBA instrument 
on the JCMT. The data set has also been analyzed by \citet{borys03,borys04}.
The mosaicked submillimeter image has non-uniform $0.4-4$~mJy point-source sensitivity 
and covers 0.034~deg$^2$ (43\% of the MIPS area) in the GOODS-N field. Forty-five $3\sigma$ 
and seventeen $4\sigma$ submillimeter sources are in the catalog of \citet{wang04}. 
We also used the submillimeter data to determine the 850~$\mu$m fluxes of the various 
samples. We measured the submillimeter fluxes and errors for these samples using optimal 
beam-weighted extractions \citep{wang04} throughout the area covered by our submillimeter 
image.

\subsection{Near-Infrared Data}

We carried out deep $J$ and $H$-band imaging of the entire GOODS-N region using the
Ultra-Low Background Camera (ULBCAM) on the University of Hawaii 2.2~m telescope during 
2004 and 2005 \citep{trouille06}. ULBCAM consists of four 2k$\times$2k HAWAII-2RG
arrays \citep{loose03} with a total $16\arcmin \times 16\arcmin$ field of view. The 
images were taken using a 13-point dither pattern with $\pm30\arcsec$ and $\pm60\arcsec$ 
dither steps in order to cover the chip gaps. The data were flattened using median sky 
flats from each dither pattern. The image distortion was corrected using the astrometry 
in the USNO-B1.0 catalog \citep{monet03}. The flattened, sky-subtracted, and warped 
images were combined to form the final mosaic with a $20\arcmin \times 20\arcmin$ area 
fully covering the GOODS-N region. The integration times at each pixel are 19 hours
in $J$ and 12.5 hours in $H$, respectively, and the $5\sigma$ sensitivities are 
0.84~$\mu$Jy and 2.06~$\mu$Jy corresponding to $5\sigma$ AB magnitude limits of 24.1 
and 23.1, respectively. A more extensive description of the data reduction and a 
detailed analysis may be found in \citet{trouille06}.

We generated our source catalogs with the SExtractor package \citep{bertin96}. Because 
the typical seeing was $0\farcs7$ and many of the sources appear extended in the images, 
we used auto aperture in SExtractor to ensure that the measured fluxes are close to the 
total fluxes. In our stacking analysis, we also consider a catalog of all $H(AB)<24$ 
(roughly the $3\sigma$ limit) sources that lie within the submillimeter image. 
Although this catalog will include a small number of false detections
($\sim1\%$, inferred from negative sources), this is not a 
significant issue for our stacking analysis, since these sources will not affect the 
signal and will only add a small amount to the noise.

\subsection{IRAC Data}

We used the GOODS-N \emph{Spitzer} Legacy Science Program first, interim, and second 
data release products (DR1, DR1+, DR2; \citealp{dickinson06}). We combined the reduced 
DR1 and DR2 IRAC superdeep images, weighted by exposure time,
to form 3.6, 4.5, 5.8, and 8.0~$\mu$m images that fully cover the GOODS-N area. 
We again generated the source catalogs in each band 
with the SExtractor package \citep{bertin96}. We detected approximately 9800, 8500, 
3600, and 3000 sources at $>5\sigma$ at 3.6, 4.5, 5.8, and 8.0$~\mu$m, respectively.  
We measured the source fluxes with fixed apertures of $4\farcs8$ (3.6 and 4.5$~\mu$m) 
and $6\arcsec$ (5.8 and 8.0$~\mu$m). These apertures are approximately three times the 
$\sim 1\farcs7$ (3.6$~\mu$m) to $\sim 2\arcsec$ (8.0$~\mu$m) FWHM of the point spread
function (PSF) and are a
good compromise between the PSF size and the source separation.  We applied aperture 
corrections from the IRAC in-flight PSFs (January 2004) to the measured fluxes.  The 
aperture corrections we used are consistent with the ones published in the IRAC Data 
Handbook. The corrected IRAC fluxes should be reasonably close to the total fluxes of 
the sources, because the majority of the sources are point-like compared to the 
$\sim2\arcsec$ IRAC PSF. The primary errors in the photometry are caused by the high 
density of sources, especially at 3.6 and 4.5$~\mu$m. In these two bands, the typical 
distance between sources is comparable to the PSF, and the maps are confusion-limited.  
Consequently, both the background estimate and the aperture photometry are highly 
subject to blending with nearby sources.

\subsection{MIPS Data}
\label{secmips}

At the longest MIR wavelengths, our sample consists of the MIPS 24$~\mu$m GOODS-N data.
We directly used the DR1+ MIPS source list and the version 0.36 MIPS map provided by the 
\emph{Spitzer} Legacy Program. The source catalog is flux-limited at 80$~\mu$Jy and 
is a subset of a more extensive catalog \citep{chary06}. With the 0.065~deg$^2$ area 
coverage, the catalog contains 1199 24$~\mu$m sources and is $>80\%$ complete at 80$~\mu$Jy
\citep{papovich04}.
The source positions are based on sources detected in the deep IRAC images, 
and the fluxes are derived using PSF fitting. The flux limit and accuracy of this catalog 
are sufficient for our purposes, so we did not attempt to generate our own MIPS source 
catalog.

Nevertheless, because the DR+ MIPS source list is still preliminary, we 
verified the sources in the list using the MIPS map. We used a 
normalized MIPS PSF to convolve with the MIPS map and measured the 24$~\mu$m 
fluxes at the cataloged positions. These PSF-weighted fluxes are 
mostly consistent with the fluxes in the DR+ source list. However, there are
a few sources at the edges of the MIPS map, where it is noisier, which have 
low-significance fluxes in our measurements. Seventeen of
these sources have no obvious counterparts in the deep IRAC, NIR, and optical
images. These sources are likely spurious. Thus, for our stacking analysis, we used 
a restricted area that was fully covered by the \emph{Hubble Space Telescope} ACS 
GOODS-N observations, since this is where nearly all of our submillimeter coverage is. 
This fully avoids the edge problems described above.

A $-0\farcs38$ offset in declination was applied to the source positions 
to match the radio-frame astrometry \citep{richards00}. 

\subsection{Optical Data}
\label{secopt}

\citet{capak04} presented ground-based deep optical imaging of  
a very wide-field region around the HDF-N. The imaging covers the 
whole MIPS and IRAC area at $U$, $B$, $V$, $R$, $I$, $z^{\prime}$, and 
$HK^{\prime}$ bands. We searched for counterparts to the various samples in 
the catalog of \citet{capak04} using a $1\arcsec$ search radius. This search 
radius closely matches the PSF in the optical and the astrometry errors in
the optical and MIR. 

Where the images overlapped, we also cross-identified the various
samples with the ACS GOODS-N catalog of \cite{giavalisco04}.

\subsection{Radio and X-ray Data}
\label{secradio}

We used the 1.4~GHz catalog and image from \citet{richards00}, which contains 
sources to a flux limit of 40$~\mu$Jy ($5\sigma$), to analyze the radio-selected 
submillimeter sample. We also used the radio data to determine the 20~cm fluxes 
of the various samples. We measured the 20~cm fluxes in $3''$ 
diameter apertures, adjusting the normalization to match the measured fluxes in 
the \citet{richards00} catalog for the overlapping set of objects. We measured 
the noise level by determining the fluxes at a large number of random positions 
and then measuring the dispersion. The $1\sigma$ noise is $14~\mu$Jy, 
which is almost a factor of two higher than that 
measured by \citet{richards00}. This reflects the large aperture used. 
However, the noise distribution measured in this way
is well fitted by a Gaussian, with an average zero flux level. 

Finally, we used the \emph{Chandra} Deep Field-North (CDF-N) 2~Ms catalog 
\citep{alexander03} to determine the X-ray properties of the various samples 
and to identify sources that contain AGNs.

\begin{deluxetable}{lccc}[!h]
\tablecaption{Optical, NIR, and MIR Photometry \label{tab1}}
\tabletypesize{\footnotesize}
\tablehead{\colhead{Band} & \colhead{Sensitivity Limit} & 
\colhead{Telescope} & \colhead{Reference} \\
&($\mu$Jy)&&}
\startdata
MIPS 24 $\mu$m 	& 80.0 	& \emph{Spitzer} & \small a \\
IRAC 3.6 $\mu$m	& 0.327	& \emph{Spitzer} & \small a \\
IRAC 4.5 $\mu$m & 0.411	& \emph{Spitzer} & \small a \\
IRAC 5.8 $\mu$m & 2.27	& \emph{Spitzer} & \small a \\
IRAC 8.0 $\mu$m & 2.15	& \emph{Spitzer} & \small a \\
$U$ 		& 0.052	& KPNO 4~m	 & \small b \\
$B$ 		& 0.063	& Subaru	 & \small b \\ 
$V$ 		& 0.069	& Subaru	 & \small b \\
$R$ 		& 0.083	& Subaru 	 & \small b \\
$I$ 		& 0.209	& Subaru	 & \small b \\
$z^{\prime}$ 	& 0.251	& Subaru	 & \small b \\
$J$ 		& 0.839	& UH 2.2~m	 & \small c \\
$H$ 		& 2.06	& UH 2.2~m	 & \small c \\
\enddata
\tablecomments{The MIPS 24~$\mu$m sample is flux-limited and complete at 80~$\mu$Jy.  
The median $1\sigma$ sensitivity of the MIPS map is 6.4~$\mu$Jy. For the
rest of the bands, the sensitivity limits are $5\sigma$ limits.
References: (a) GOODS \emph{Spitzer} Legacy Program DR1, DR1+, and DR2; (b)
\citet{capak04}; (c) this work. }
\end{deluxetable}

\begin{figure}[!h]
\epsscale{1.0}
\plotone{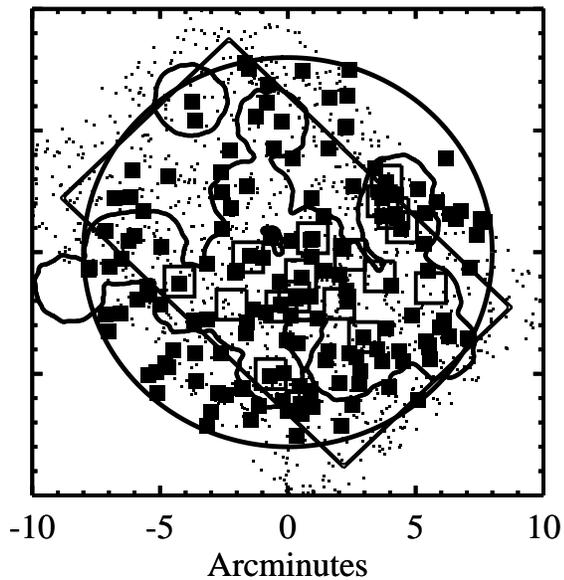}
\caption{Schematic layout of the various images. The rectangle shows the deepest portion 
of the ACS GOODS-N image. The contours show the SCUBA image. The circle shows an 
$8'$ radius around the radio and X-ray centers. The large open squares denote
the 17 SCUBA sources detected at the $4\sigma$ level, the solid squares denote the 20~cm 
sources within the $8'$ radius, and the dots denote the $24~\mu$m sample. 
The total submillimeter area is 125~arcmin$^{2}$, and the area of overlap with the ACS GOODS-N 
rectangle, which we use as our primary area, is 106~arcmin$^{2}$. The ACS GOODS-N rectangle
has an area of 144~arcmin$^{2}$.} 
\label{schema}
\end{figure}

\subsection{Data Summary}

We summarize the flux limits of the various samples in Table~\ref{tab1}. We also show  
the geometry of the various data sets schematically in Figure~\ref{schema}. The 
ground-based optical and NIR images cover the entire area shown. We mark 
the region with complete coverage from the ACS GOODS-N data with a rectangle and the 
submillimeter region with the contours. The overlap area constitutes our core area
and covers 106~arcmin$^{2}$. Essentially all of this region lies within 
$8'$ \emph{(circle)} of the X-ray and radio centers, where the X-ray and radio images 
have relatively uniform sensitivity. We denote the $4\sigma$ submillimeter sources with 
large open squares, the 20~cm sources with smaller solid squares, and the 24$~\mu$m 
sources with dots. Our core area excludes the more poorly sampled regions of the 
24$~\mu$m image.

\section{Redshifts}
\label{secz}

\subsection{Spectroscopic Redshifts}
\label{secspz}

Intensive spectroscopic redshift surveys have been carried out 
in the ACS GOODS-N region. We searched for spectroscopic redshifts for 
the sources in \citet{wirth04}, \citet{cowie04}, and \citet{chapman05}.  A
substantial number of additional redshifts, which either lie outside
the ACS GOODS-N region or were obtained from our spectroscopic
runs with the Deep Extragalactic Imaging Multi-Object Spectrograph
(DEIMOS; \citealt{faber03}) on the Keck 10~m telescope subsequent
to the publication of these papers, were also included.

\subsection{Photometric Redshifts}
\label{secphotz}

We used the $U$ to 8~$\mu$m photometry of the sources to derive their 
photometric redshifts. We only considered sources that were detected 
in at least five bands. Compared to most optical photometric redshifts, 
adding the deep $J$ and $H$ magnitudes and the \emph{Spitzer}  
data has the advantage of improving the high-redshift end of the photometric 
redshift determinations. The MIR photometry probes various 
spectral features, including the 1.6$~\mu$m bump caused by the opacity 
minimum in the stellar atmosphere and PAH emission at $5-9~\mu$m. 
In fact, the \emph{Spitzer} photometry alone has been used to derive photometric 
redshifts based on the 1.6$~\mu$m bump \citep[see e.g.,][]{sawicki02,egami04}.

\begin{figure}[!h]
\epsscale{0.9}
\plotone{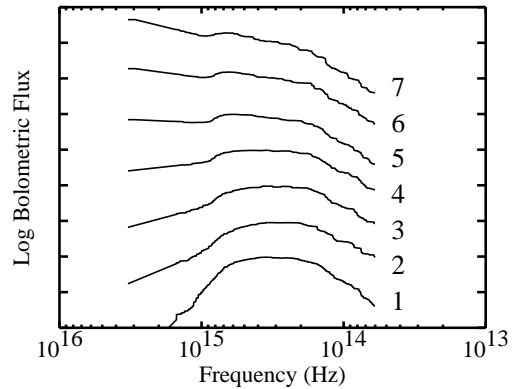}
\caption{SED templates used in the training-set photometric redshift determinations.  
Vertical axes are arbitrarily scaled bolometric fluxes ($\nu F_{\nu}$). The numerical
labels give the SED classes, which range from class 1 (elliptical galaxies) to class 7
(very blue star-forming galaxies). 
\label{figtemp}}
\end{figure}

\begin{deluxetable*}{lrrrrrrr}[!h]
\tablecaption{Properties of the $850~\mu$m Sources with 20~cm and $24~\mu$m Matches \label{tab2}}
\tabletypesize{\small}
\tablehead{\colhead{ID}  &
\colhead{$z_{sp}$} &
\colhead{$z_{ph}$} &
\colhead{$S_{850\mu\rm m}$} &
\colhead{$S_{1.4 \rm GHz}$} &
\colhead{$S_{3.6\mu\rm  m}$} &
\colhead{$S_{24\mu\rm  m}$} &
\colhead{Log($S_{\rm HX}$)} \\
&\colhead{} &
\colhead{} &
\colhead{(mJy)} &
\colhead{($\mu$Jy)} &
\colhead{($\mu$Jy)} &
\colhead{($\mu$Jy)} &
\colhead{(ergs~cm$^{-2}$~s$^{-1}$)}}
\startdata
GOODS 850-3  & 1.865 & 2.71 & $ 7.7\pm1.0$ & $151$   &  14.4 & $ 330$ & \nodata \\
GOODS 850-6  & \nodata & 2.17 & $13.6\pm2.3$ & $107$ &   9.5 & $ 185$ & \nodata \\
GOODS 850-7  & 2.578 & 3.75 & $ 6.2\pm1.0$ & $ 53.9$  &  9.8 & $ 313$ & $-15.10$ \\
GOODS 850-9  & 2.490 & 2.31 & $ 7.1\pm1.2$ & $ 45.3$ &  16.0 & $ 235$ & $-15.04$ \\
GOODS 850-11 & \nodata & 2.67 & $10.8\pm2.2$ & $124$ &   4.4 & $ 165$ & $<-15.85$ \\
GOODS 850-15 & \nodata & 2.91 & $ 8.7\pm2.0$ & $148$ &  13.4 & $ 370$ & \nodata \\
GOODS 850-16 & \nodata & 3.79 & $12.4\pm2.9$ & $324$ &  15.4 & $ 267$ & \nodata \\
GOODS 850-17 & 1.013 & 1.22 & $ 5.7\pm1.4$ & $ 81.4$ &  86.7 &  $ 724$ & $-14.65$ \\
\enddata
\end{deluxetable*}

The primary difficulty of incorporating the MIR data into the photometric 
redshift estimation is the lack of optical to MIR galaxy spectrum templates. 
\citet{perez05} overcame this problem by building ``training-set''
spectral energy distributions (SEDs) from 
the sources within the MIPS sample itself. We also used this method to generate 
our photometric redshifts. We used just over 1200 galaxies with known redshifts 
and spectral types in the ACS GOODS-N region to construct seven templates over 
the frequency range from $6\times10^{13}$~Hz to $4\times10^{15}$~Hz.
These templates range
from an elliptical galaxy spectrum to a very blue star-forming galaxy spectrum. 
The seven templates are shown in Figure~\ref{figtemp}. We then made a least-squares 
fit to these templates to determine the photometric redshifts and spectral types
for the galaxies in each sample.

\begin{figure}[!h]
\epsscale{0.9}
\plotone{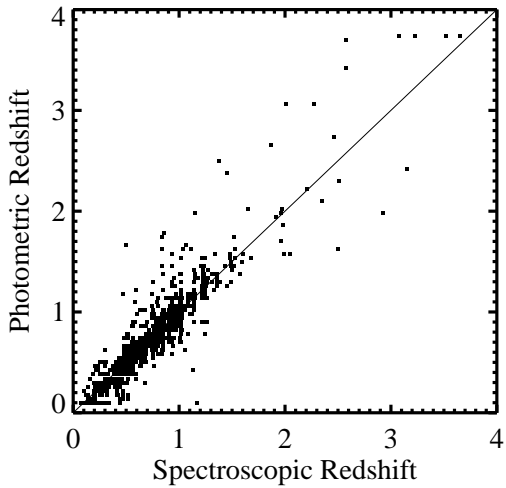}
\caption{Comparison of our photometric redshifts, as determined from the training-set 
templates, with the spectroscopic redshifts for the spectroscopically-identified $H$-band 
sample in the ACS GOODS-N region.
\label{figphotz}}
\end{figure}

The method works extremely well over a wide range of redshifts and only fails 
for a small number of sources. 
In Figure~\ref{figphotz}, we compare our photometric redshifts with the spectroscopic 
redshifts for the spectroscopically-identified $H<23$ sample in the ACS GOODS-N region. 
There are 1213 sources in this sample with spectroscopic redshifts that
are not saturated in the optical ($z'>19$), and 1134 of these have statistically
acceptable
fits to the templates at some redshift from $z=0$ to 4. The remaining sources
either have unusual SEDs or are blended with a neighbor. 
For the 1134 sources with both spectroscopic and photometric redshifts shown
in Figure~\ref{figphotz},
there are only a couple of seriously discrepant sources, and while 
the scatter becomes larger at the higher redshifts, the method 
robustly places nearly all the sources in the correct redshift range.

\section{Identification of the 850 micron EBL}
\label{secebl}

\subsection{Direct Identification of Bright Submillimeter Sources}

Many of the bright ($>2$~mJy) submillimeter sources that are directly detected
in the $850~\mu$m blank-field SCUBA images can be localized by their radio emission,
as discussed in the introduction. However, since the $24~\mu$m-band traces
the short wavelength end of the FIR emission, the $24~\mu$m sample may provide 
additional counterparts to the bright SCUBA sources \citep{egami04,ivison04}.

Our $850~\mu$m SCUBA map covers 0.034~deg$^2$, and there are 17 $4\sigma$ bright 
SCUBA sources in this area in the catalog of \cite{wang04}, whose notation 
we shall follow. Three of the bright SCUBA sources (GOODS 850-2, 850-8, and 850-14) 
have no $24~\mu$m counterparts, even within a very wide $8\arcsec$ search radius.
These sources also do not have radio counterparts \citep{wang04}. The remaining 14 
have a total of twelve $24~\mu$m sources within a $6\arcsec$ search radius and
eighteen $24~\mu$m sources within an $8\arcsec$ search radius. Given the number 
density of $24~\mu$m sources, we expect $\sim3$ false matches at $6\arcsec$ and 
$\sim5$ false matches at $8\arcsec$, so we are identifying some $9-13$ of the 
remaining 14 sources, or about $50-75$\% of the total sample.

We can compare this with the counterpart identifications made using the 20~cm data
by \citet{wang04}. Eight of the 17 $4\sigma$ bright SCUBA sources have 20~cm counterparts
within $6\arcsec$, and we expect all of these to be real, given the surface density
of radio sources. Ten have counterparts within $8\arcsec$, including one source
with two radio counterparts; we expect one false match at this radius. These results
suggest that the $24~\mu$m data may be slightly better at picking up the 
bright SCUBA sources, but the difference is not statistically significant.
All eight radio sources with bright SCUBA counterparts within $6\arcsec$ are seen 
at $24~\mu$m, and the $24~\mu$m data then picks out a further four sources, of which
three are expected, on average, to be false.

The redshifts (spectroscopic and photometric) and properties of our eight securely 
identified bright SCUBA sources are summarized in Table~\ref{tab2}.
The whole $24~\mu$m sample has a median $24~\mu$m flux of 152~$\mu$Jy. The SCUBA 
sources in Table~\ref{tab2} have a median $24~\mu$m flux of 300~$\mu$Jy and 
are weighted to the high end of the $24~\mu$m fluxes. They are mostly at high 
redshifts ($z=2-3$), with a median redshift of $z=2.5$, and are consistent with 
being ULIRGs. \citet{ivison02} presented a sample
of 30 radio-identified SCUBA sources with $850~\mu$m fluxes $>8$~mJy.
Using the millimetric redshift technique, they found a median redshift of $z=2.4$.
\citet{chapman03a} presented a spectroscopic redshift sample of 10 radio-identified 
bright ($>2$~mJy) SCUBA sources, and \citet{chapman05} expanded that sample 
to 73 sources. Their typical redshift range is $z=1.7-2.8$, with a median 
redshift of $z=2.2$. The redshift distribution of our radio-identified bright 
SCUBA sample is fully consistent with the results of the above groups.

\begin{deluxetable*}{lccccccc}[!h]
\tablecaption{850~$\mu$m Stacking Analyses\label{tab3}}
\tablehead{\colhead{Waveband} & \colhead{Flux$_{min}$}
& \colhead{$N$} & \colhead{Area} 
& \colhead{$<S_{850}>$} & \colhead{$I_{\nu}$} & \colhead{$I_{\nu, \rm clean}$} \\
& \colhead{($\mu$Jy)}  & & \colhead{(arcmin$^{2}$)} & \colhead{(mJy)} & \colhead{(Jy deg$^{-2}$)} & \colhead{(Jy deg$^{-2}$)}}
\startdata
ACS 0.45 $\mu$m & 0.05  & 6868 & 106 & $0.10\pm0.017$ & $24.8(15.2)\pm4.1(6.0)$ & $14.3(19.3)\pm4.0(6.0)$ \\
ACS 0.8 $\mu$m & 0.1  & 7826 & 106 & $0.10\pm0.016$ & $26.8(16.3)\pm4.3(6.4)$ & $15.4(20.9)\pm4.2(6.3)$ \\
ULB 1.6 $\mu$m & 1 & 3094 & 122 & $0.20\pm0.03$ & $18.8(21.1)\pm2.3(3.5)$ & $10.4(17.7)\pm2.3(3.5)$ \\
IRAC 3.6 $\mu$m & 0.3 & 5245  & 106 & $0.11\pm0.0066$ & $19.6(16.7)\pm3.4(5.1)$ & $8.2(15.6)\pm3.4(5.1)$ \\
IRAC 8.0 $\mu$m & 2.0 & 1587  & 106 & $0.33\pm0.03$ & $18.0(19.0)\pm1.9(2.8)$ & $9.1(13.8)\pm1.9(2.8)$ \\
MIPS 24 $\mu$m & 80 & 493 & 106  & $0.66\pm0.06$ & $11.4(9.1)\pm1.1(1.5)$ & $5.9(5.6)\pm1.1(1.6)$ \\
VLA  20 cm & 40& 101 & 122 & $1.31\pm0.13$ & $4.0(5.1)\pm0.40(0.6)$ & $1.5(2.7)\pm0.4(0.6)$ \\
NIR 1.6, 3.6 $\mu$m & 1.8 & 3121 & 106 & $0.20\pm0.025$ & $20.7(20.9)\pm2.6(4.6)$ & $11.4(16.1)\pm2.6(4.0)$ \\
\enddata
\tablecomments{The $N$ values are the numbers of sources with errors less than 
4~mJy that were used in the stacking analyses.}
\end{deluxetable*}

\subsection{A Stacking Analysis of the Submillimeter EBL}

We cannot directly analyze the submillimeter sources that are too faint to be detected 
in blank-field SCUBA surveys, but we can use our galaxy samples at other wavelengths to 
study these sources using a stacking technique. This technique was first used by 
\cite{peacock00} to study the submillimeter properties of Lyman-break galaxies.  Here we 
use our SCUBA map to determine the average 850~$\mu$m properties of the sources in a 
given sample, and hence the amount of 850~$\mu$m light they produce. Although such an 
analysis provides very little information on the individual source properties, it can 
show which of the galaxy populations give rise to the submillimeter light. Since we can 
also break the samples down by galaxy flux, color, spectral type, or redshift, we can 
determine the properties and redshift distribution of the class(es) of sources 
that is (are) producing the bulk of the 850~$\mu$m EBL.
 
One difficulty with trying to assess what fraction of the $850~\mu$m EBL a class of sources 
produces is that the absolute normalization of the $850~\mu$m EBL is somewhat uncertain.
The observed $850~\mu$m EBL is given as 31~Jy~deg$^{-2}$ in \citet{puget96}
and as 44~Jy~deg$^{-2}$ in \citet{fixsen98}. The substantial difference between the two 
estimates is a consequence of different corrections for foreground emission, and it is 
unclear which is the better estimate. We therefore compare with the full range in our 
subsequent analysis.

The possible source populations that are contributing the bulk of the 850$~\mu$m
EBL are also constrained by the deepest submillimeter number counts. The  
$850~\mu$m number counts of \citet{cowie02} showed that the number of submillimeter 
sources with fluxes $>0.5$~mJy is approximately $2.5\times10^4$~deg$^{-2}$.  
This is therefore a rough minimum on the density of any sample that seeks
to explain the $850~\mu$m EBL. Simply based on such number densities,
it is clear that there are not enough 20~cm selected sources to account
for the light, and there are only marginally enough $24~\mu$m sources
(approximately $1.8\times10^4$~deg$^{-2}$ above $80~\mu$Jy). In other words,
the current 20~cm and $24~\mu$m samples are not deep enough to have reached the fainter 
submillimeter sources, while the NIR and MIR samples are deep enough to have surface 
densities that are substantially above the required value.

We first computed the contributions to the $850~\mu$m EBL from seven samples: the 
$H$-band, 3.6$~\mu$m, 8$~\mu$m, $24~\mu$m, and 20~cm selected samples, as well as the 
$B$ and $I$-band selected samples from the ACS GOODS-N catalogs of \cite{giavalisco04}. 
For each source in each sample, we determined the beam-weighted $850~\mu$m flux and 
noise from our SCUBA map (see \S\ref{secsmm}). 
We then measured the error-weighted average $850~\mu$m flux for all of the sources 
in the sample that had errors less than the chosen cut value of 4~mJy. 
We also determined the area where each sample had submillimeter measurements
of this sensitivity. The EBL contribution is then given by the
product of the number of sources and their error-weighted mean
divided by the observed area. The results are almost identical for cut values 
other than 4~mJy. The contribution of each population to the $850~\mu$m EBL is 
summarized in Table~\ref{tab3}, where we give the sample wavelength and limiting 
flux, the area covered, the number of sources, the error-weighted mean 850~$\mu$m 
flux for each source, the $850~\mu$m EBL contribution with its $1\sigma$ error, and
the $850~\mu$m EBL contribution measured from a map where $4\sigma$ $850~\mu$m
sources are CLEANed (see below). Since the error weighting may result in a very small 
region dominating the signal, we also computed the unweighted 850~$\mu$m signal over the 
same area. This is given in parentheses in the final two columns of the table.  

Our results confirm the MIR detections made by
\citet{serjeant04} with the \emph{Spitzer} Early Release Observations and the
SCUBA 8~mJy survey, but at substantially higher significance levels. 
As an example, there are 493 $24~\mu$m sources with $850~\mu$m errors less than 4~mJy in the
SCUBA map. The mean $850~\mu$m flux of these sources is $0.66\pm0.06$~mJy. This is a 
substantial improvement in signal to noise over the value of $0.30\pm0.24$~mJy 
found by \citet{serjeant04} due to the large number of $24~\mu$m sources in the GOODS-N field.   
We obtain a consistent average flux of $0.53\pm0.08$~mJy from the unweighted sample,
which shows that the lower error regions are not dominating the signal.
The total stacking contribution of the $24~\mu$m sources to the $850~\mu$m EBL is 
$11.4\pm1.1$~Jy~deg$^{-2}$, or about $25-35$\% of the total $850~\mu$m EBL. 

As expected, given that they do not have sufficiently high number
densities, the 20~cm and $24~\mu$m samples only identify a fraction of the $850~\mu$m 
EBL. (The radio sample identifies $4.0\pm 0.40$~Jy~deg$^{-2}$, or about $10-13$\% 
of the 850~$\mu$m EBL.) It is clear that much deeper samples would be required at both 
of these wavelengths to substantially identify the $850~\mu$m EBL. To try to see how
much fainter we would have to go at $24~\mu$m, we generated a fainter $24~\mu$m source 
catalog using SExtractor and excluded the original 1173 MIPS sources. There are 731 
$3\sigma$ MIPS sources in this catalog, and 338 of them are in the SCUBA map.
Their median $24~\mu$m flux is 46~$\mu$Jy, and the faintest ones are $\sim10$~$\mu$Jy.  
These faint $24~\mu$m sources possess a weak $850~\mu$m stacking signal of 
$1.73\pm1.32$~Jy~deg$^{-2}$. We note that the completeness of the $24~\mu$m catalog 
decreases rapidly from $\sim1.0$ at 80~$\mu$Jy to $\sim0.1$ at 40~$\mu$Jy 
\citep[e.g.,][]{papovich04}. Therefore, the faint $24~\mu$m sources in the $10-80$~$\mu$Jy 
range might make a significant contribution ($>10$~Jy~deg$^{-2}$) to the 850~$\mu$m EBL.  
However, because of the low statistical significance of the stacking signal, we cannot show
that this is the case with the current 
data. To provide a definitive answer, we would need a 
substantially complete $24~\mu$m catalog to about a 10~$\mu$Jy flux.

By contrast, the 8~$\mu$m sample provides a much more substantial identification 
of the 850~$\mu$m EBL ($18.0\pm 1.9$~Jy~deg$^{-2}$), as do all of the shorter wavelength bands. 
In fact, at first sight, the optical wavelength bands give the most complete identification
of all. However, on closer inspection, much of this signal comes from the very small 
HDF-N region, where the submillimeter errors are very low. Here the slightly smaller 
EBL determined from the unweighted average may be a more representative value.

The signal from each of the samples is, of course, heavily overlapped, and we next 
compared the samples to see how much additional signal each added. To do this,
we took each sample in turn as our primary sample and then 
measured signals from the residual sources in each of the other samples after each 
of the primary sample's overlapping sources were excluded. 
That is, we measured the signal from the other samples, after excluding all sources which
were already present in the primary sample.
We were able to determine which of the samples 
were most complementary based on this procedure. As a result of our analysis, we 
formed a sample from the combination of all sources with $H$-band
or $3.6~\mu$m fluxes greater than 1.8~$\mu$Jy. We restricted the sample to the ACS GOODS-N 
region. We refer to this sample as our NIR sample, and we summarize its properties
in the final line of Table~\ref{tab3}. Physically, since the two bands bracket the peak 
flux in most galaxy SEDs over roughly the $z=0-3$ range, this sample is choosing nearly 
all of the galaxies with peak observed fluxes above the 1.8~$\mu$Jy cut.
 
From our NIR sample, we find a contribution of $20.7\pm2.6$~Jy~deg$^{-2}$,
or $50-70$\% of the $850~\mu$m EBL. Perhaps even more importantly,
our NIR sample appears to contain nearly all of the EBL that can be measured from
the remaining samples. In other words, when our NIR sample is excluded, none of 
the residual $8~\mu$m, $24~\mu$m, or radio sample sources gives a signal greater than 
1~Jy~deg$^{-2}$. The residual $B$ and $I$-band sample sources give a signal of a few 
Jy~deg$^{-2}$ in the weighted samples but a null signal in the unweighted samples.
We therefore adopt our NIR sample as our primary sample for further analysis.

There are 3121 sources in our NIR sample. Of these, 2415 are in the original $H<24$  
$3\sigma$ catalog, and a further 374 have $H$-band magnitudes brighter than 24,
reflecting the incompleteness of this catalog in the $H=23-24$ range.
The remaining 332 
sources in our NIR sample would not have been selected in a complete $H=24$ sample 
and have only been detected using the 3.6~$\mu$m catalog; they give 
a signal of $4.8\pm0.9$~Jy~deg$^{-2}$. Thus, the combined catalog gives an
improved identification of the EBL relative to either catalog alone. 

We visually inspected the $3.6~\mu$m image for all of the sources without optical
or NIR counterparts to check that these were not spurious. About 30 sources are suspect, 
either because they are not clearly seen in the $3.6~\mu$m image, or because
they are contaminated by a neighbor. However, excluding these sources has no effect 
on the measured signal, and we conclude that there is a significant contribution from 
the $3.6~\mu$m sources in the sample.

To check the robustness of our results, we performed a number of tests. 
First, we summed the total SCUBA image to find the 
total flux of our SCUBA map and found the result to be strictly zero. 
This is as expected, since each positive source in the SCUBA map has two negative 50\% 
sidelobes. The zero sum of our SCUBA map indicates that there was good sky subtraction 
during the data reduction.  
Because of the zero sum of the submillimeter map, any random populations will have zero 
stacking fluxes, and a positive stacking signal will indicate a correlation between that 
population and the submillimeter sources. To test this and the assigned statistical 
errors, we measured fluxes for large numbers of random positions in the SCUBA image
and analyzed these in the same fashion as the real samples.
The results were fully consistent with expectations in both the average
signal and the statistical spread.

The third test was a Monte Carlo 
simulation. We used the $24~\mu$m source catalog and randomized its astrometry to create 
simulated $850~\mu$m sources. The $850~\mu$m fluxes of the simulated sources were derived 
from conversions based on M82 and Arp~220. The simulated sources were added into 
the ``true noise'' map of \citet{wang04}, which has both the bright SCUBA sources and the
faint confusion sources removed, such that only noise is left. 
When we measured the $850~\mu$m fluxes from the simulated maps, we randomly offset the
measured positions from the source positions  
with up to $2\arcsec$ rms position errors, enough to account for the pointing errors 
of the submillimeter telescope and the astrometry errors in the $24~\mu$m catalog. The 
average stacking flux was measured in 100 such realizations using the
same methodology as was used in analyzing the true submillimeter
image and fully recovered the known input fluxes. 
In summary, our Monte Carlo simulations and the zero sum of the SCUBA map show that 
the stacking flux is an unbiased estimate of the 850~$\mu$m EBL and that the assigned 
errors are realistic.

As our fourth test, we split the NIR sample into two parts: one corresponding
to sources where the measured submillimeter errors lay between 0.1 and 1.5~mJy,
and the other to sources where the errors lay between 1.5 and 4~mJy. The first sample 
contained 826 sources and gave an 850~$\mu$m EBL contribution of $20.1\pm3.2$~Jy~deg$^{-2}$. 
The second sample contained 2295 sources and gave an 850~$\mu$m EBL contribution of 
$21.8\pm4.5$~Jy~deg$^{-2}$, showing that an equivalent signal can be obtained from 
two samples with very different sensitivities.

Correlations in the target sample can result in
an upward bias to the signal if the submillimeter flux comes from
overdense regions, since, in this case, we overcount the mean flux associated
with each source (e.g., \citealt{serjeant03}). Given the expected angular correlations
in the samples, we do not expect this effect to be large, but it may be
present at some level. We tested this by measuring the average number of galaxies
lying within $7"$ of a given galaxy. Since this is the half-width of the $850~\mu$m
beam, we expect that only galaxies within this area could be substantially
contaminated. For the NIR sample described above, which has a surface density
of $1.1\times10^{5}$ deg$^{-2}$, we found an average number of galaxies
of 1.1 within this radius. This is actually smaller than the expected number
for a random distribution (1.18), so there is very little correlated signal,
and the correlation effects on the stacking signal are small. Since we shall subsequently
restrict the sample even further, we ran this same test on that restricted sample
and found a similar result. 

Other groups (e.g., \citealp{serjeant04}) perform stacking analyses on 
maps where known submillimeter sources are removed.  As our final test, we also 
removed all known submillimeter sources detected at or above the $4~\sigma$ level 
from the map and reran our stacking analyses.  To do this we subtracted (CLEANed) 
the fitted PSFs of all of these sources from the image.  Because our noise estimate 
does not include the confusion noise (e.g., Cowie et al.\ 2002; Wang et al. 2004), our 
$4~\sigma$ cut is comparable to the $3.5~\sigma$ cut used by most other groups. 
We note that such a sigma cut has different flux limits in different areas of
our map.  The totol flux removed from the map is only 5.6 Jy deg$^{-2}$ but the 
sources removed from the deepest areas indeed correspond to a total surface 
brightness of $\sim10$ Jy deg$^{-2}$ according to the number counts in Wang et al.\ (2004).  
The removal therefore affects the weighted contribution to the EBL more than
the unweighted contribition and the later value is comparable to the reduction in the 
weighted signal while the former matches the change in the unweighted signal.  
As a consequence of this complex biasing effect from the non-uniform sensitivities, 
we decided to use only our unCLEANed results.  However, for completeness,
we also include our CLEANed results in Table~3.

\subsection{The $850~\mu$m EBL versus Galaxy Flux, Color, and Spectral Type}

We can further subdivide the contributions to the 850$~\mu$m EBL by galaxy flux,
color, and spectral type to determine the properties of the sources giving rise 
to the light.

\begin{figure}[!h]
\epsscale{1.14}
\vskip 0.15cm
\plotone{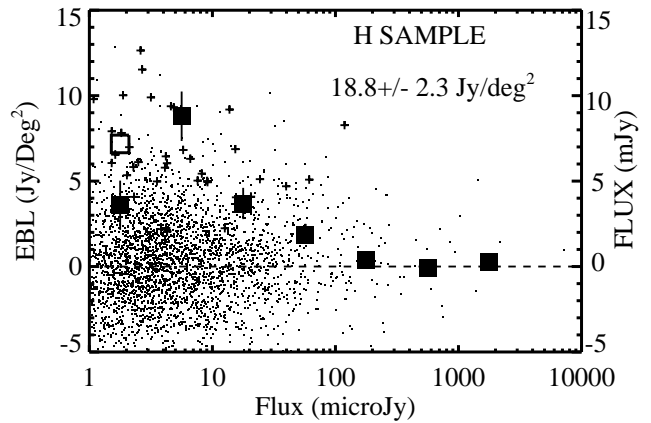}
\caption{Contributions to the $850~\mu$m EBL from the $H$-band sample vs. the
$H$-band flux. The solid squares show the contributions from each half dex flux 
interval with $1\sigma$ error bars. The number in the upper right corner gives
the total contribution. The open square shows the contribution from the lowest
flux interval corrected for the incompleteness in the $H$-band catalog. 
The dots denote the measured fluxes of the individual sources for a y-axis 
in mJy units, and the plus signs denote sources with $4\sigma$ measurements
at 850~$\mu$m. \vskip 0.05cm
\label{ebl_flux}}
\end{figure}

\begin{figure}[!h]

\epsscale{1.14}
\plotone{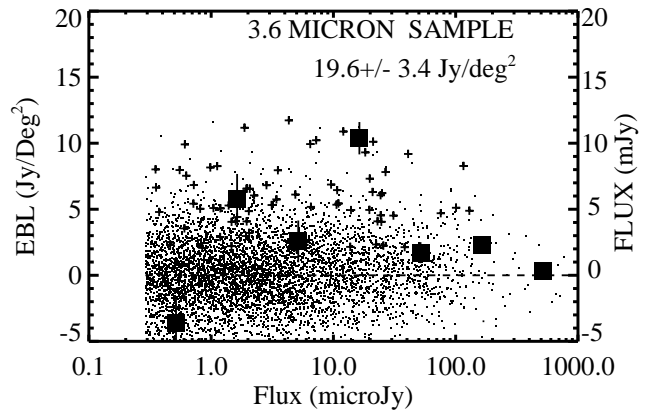}
\caption{Contributions to the $850~\mu$m EBL from the $3.6~\mu$m sample vs. the
$3.6~\mu$m flux. The solid squares show the contributions from each half dex flux 
interval with $1\sigma$ error bars. The number in the upper right corner gives the 
total contribution. The dots denote the measured fluxes of the individual sources 
for a y-axis in mJy units, and the plus signs denote sources with 
$4\sigma$ measurements at 850~$\mu$m. \vskip 0.05cm
\label{ebl_flux2}}
\end{figure}

We show the contributions to the $850~\mu$m EBL from the $H$-band sample versus 
the $H$-band flux in Figure~\ref{ebl_flux} and from the $3.6~\mu$m sample versus 
the $3.6~\mu$m flux in Figure~\ref{ebl_flux2}. In each case, we denote with
solid squares the contributions in half dex flux intervals, starting at the 
limiting flux of the sample. The total contribution
of the entire sample is given in the upper right corner.
The individual source fluxes in mJy are denoted by dots; $4\sigma$ measurements 
are denoted by plus signs. For the lowest flux interval in the $H$-band sample, we 
also show the EBL value corrected for the incompleteness in 
the $H$-band catalog \emph{(open square)}.
For both samples, the contributions are dropping at the faintest end, which would 
be consistent with the onset of convergence to the asymptotic value. The peak 
contribution comes near a flux of 6~$\mu$Jy in the $H$-band and near a flux of 
18~$\mu$Jy in the $3.6~\mu$m-band. 
Expressed in magnitudes, the peak contribution to the measured 
signal is coming from sources with NIR magnitudes of $21-22$ (AB).

\begin{figure}[ht]
\epsscale{1.05}
\plotone{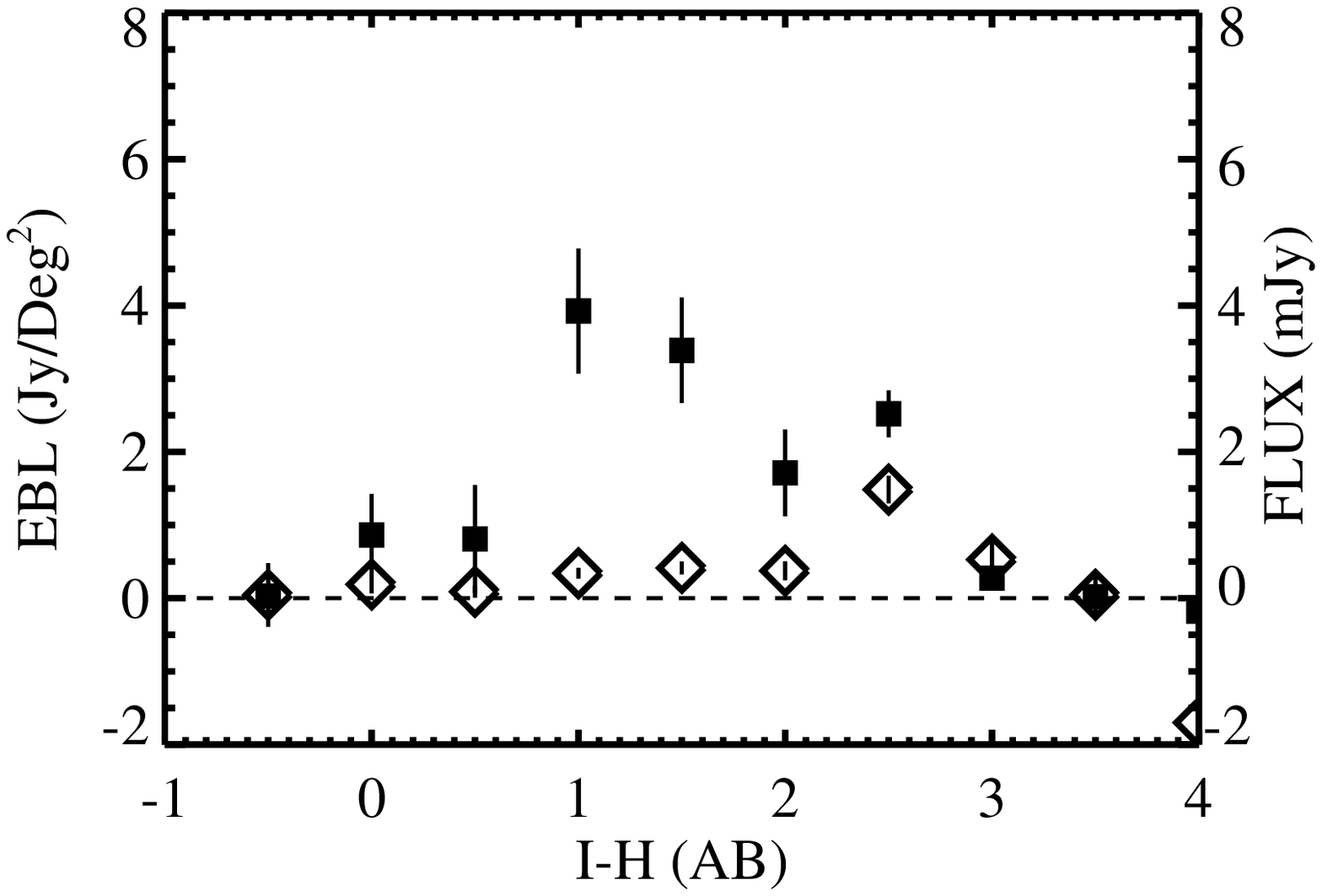}
\caption{Contributions to the $850~\mu$m EBL from the $H$-band sample
vs. the $I-H$ color in AB magnitudes. The solid squares show the contributions 
from each color interval with $1\sigma$ error bars. The open diamonds show the
mean $850~\mu$m fluxes of the sources for a y-axis in mJy units.
\label{hm_color}}
\end{figure}

\begin{figure}[ht]
\epsscale{1.05}
\plotone{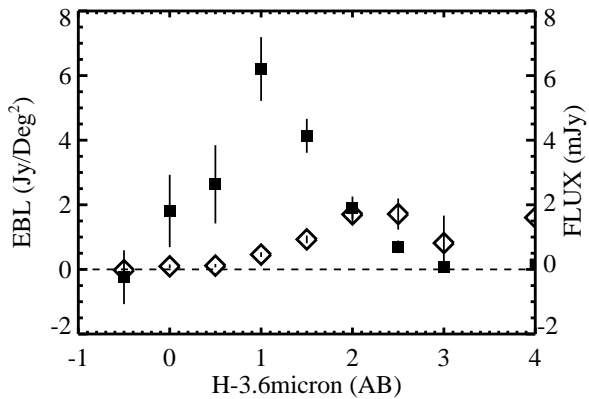}
\caption{
Contributions to the $850~\mu$m EBL from the 3.6~$\mu$m sample
vs. the $H-3.6~\mu$m color in AB magnitudes. The solid squares show the 
contributions from each color interval with $1\sigma$ error bars. 
The open diamonds show the mean 850~$\mu$m fluxes of the sources for a 
y-axis in mJy units.
\label{m1_color}}
\end{figure}

We show the contributions to the $850~\mu$m EBL from the $H$-band sample versus 
the $I-H$ color in Figure~\ref{hm_color} and from the 3.6~$\mu$m sample versus 
the $H-3.6~\mu$m color in Figure~\ref{m1_color}. In both cases, the light comes 
from the redder sources in the sample. 
In the $H$-band sample, the $I-H$ color weighted by the submillimeter light 
contribution is 1.5, while the mean color of the whole sample is 1.0. This arises 
because there is a higher mean submillimeter signal per object \emph{(open diamonds)} 
in the red-colored sources. Similarly, the $3.6~\mu$m sample has a 
submillimeter-weighted color $H-3.6~\mu$m of 1.2, as compared to a mean color of 
the whole sample of $-0.3$. The cumulative signal in the $H$-band sample is 
$3.4\pm1.1$~Jy~deg$^{-2}$ for $I-H>2$ and $14.5\pm2.1$~Jy~deg$^{-2}$ for $I-H>1$.

This biased contribution of red galaxies to the submillimeter light is a known result 
from stacking analyses carried out using submillimeter measurements of extremely
red objects in both lensed cluster and blank-field surveys 
\citep{wehner02,webb04,knudsen05}.  

\begin{figure}[!h]
\epsscale{1.05}
\plotone{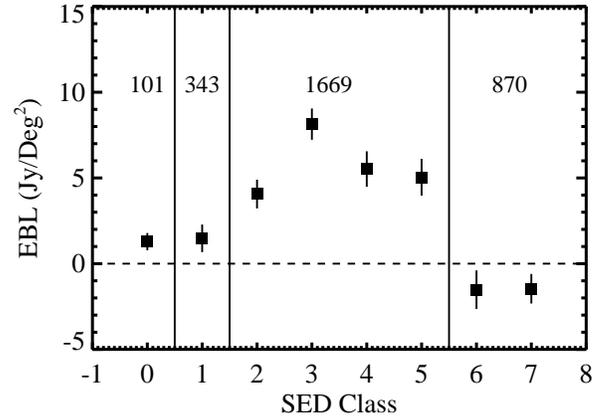}
\caption{
Contributions to the 850~$\mu$m EBL vs. the SED class. Unidentified sources are 
placed in class 0, and spectroscopically-identified stars are excluded. 
The solid squares show the contributions from each class with $1\sigma$ error bars. 
The vertical lines divide the sources into 4 bins, with the number of sources
in each bin printed at the top of that bin. 
\label{smmclass}}
\end{figure}

However, the observed-frame colors are a function of galaxy type, reddening,
and redshift, and we may make a much more powerful analysis using the
photometric redshift determinations, which separately yield the redshift
and the rest-frame SED for each source. In Figure~\ref{smmclass}, we show the 
850~$\mu$m EBL contribution from the galaxies in the NIR sample divided into the 
seven SED classes used in our training-set photometric redshift analysis 
(see Fig.~\ref{figtemp}). We assign 
unidentified sources an SED class of zero, and we exclude spectroscopically-identified 
stars. The vertical lines split the sample into four bins: unidentified sources (class 0), 
elliptical galaxies (class 1), intermediate spiral galaxies (classes $2-5$), and very 
blue star-forming galaxies (classes 6 and 7). We have printed directly on the figure 
the number of sources in each of the four bins.

\begin{figure}[!h]
\epsscale{1.10}
\plotone{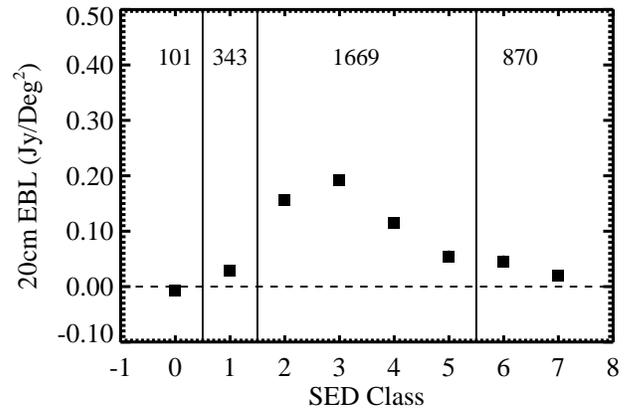}
\caption{
Contributions to the 20~cm EBL vs. the SED class. Unidentified sources are placed
in class 0, and spectroscopically-identified stars are excluded. The solid
squares show the contributions from each class with $1\sigma$ error bars. 
(These are smaller than the symbol size.) The vertical lines divide the sources 
into 4 bins, with the number of sources in each bin printed at the top of that bin.
\label{frad_class}}
\end{figure}

\begin{figure}[!h]
\epsscale{1.05}
\plotone{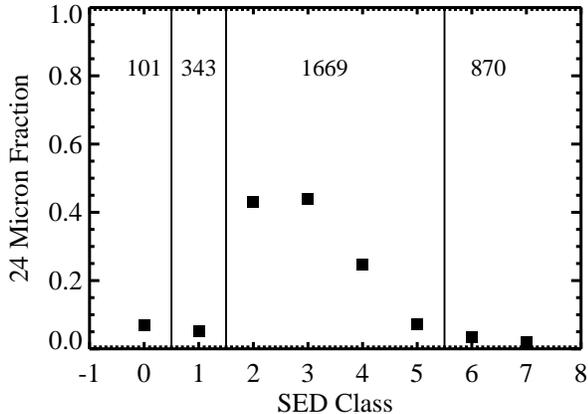}
\caption{
Fraction of the sources detected above a 24~$\mu$m flux of 80~$\mu$Jy vs. the SED class.
Unidentified sources are placed in class 0, and spectroscopically-identified stars are
excluded. The vertical lines divide the sources into 4 bins, with the number of sources
in each bin printed at the top of that bin.
\label{f24_fract}}
\end{figure}

Neither the elliptical galaxies nor the very blue star-forming galaxies give a significant 
signal. The 1213 galaxies in these three classes have an error-weighted 850~$\mu$m flux of 
$-0.03\pm0.04$~Jy, presumably reflecting the lack of star formation in the elliptical 
galaxies and the absence of dust in the blue galaxies. Nearly all of the 
850~$\mu$m signal comes from the intermediate spiral galaxies, which have an
error-weighted 850~$\mu$m flux of $0.40\pm0.03$~Jy. There is a small contribution 
to the 850~$\mu$m EBL from the unidentified sources, which have an error-weighted 
850~$\mu$m flux of $0.37\pm0.14$~Jy. 

The same selection appears if we consider the 20~cm or 24~$\mu$m properties of the 
NIR sample as a function of SED class. In order to determine the average 
20~cm flux of the NIR sample, we performed a stacking 
analysis using the 20~cm image of \citet{richards00}. (For the radio
image, since the errors are constant, the average signal is the same
as the error-weighted signal.) As with the submillimeter analysis, 
we measured the 20~cm fluxes for each of the sources in the NIR sample 
(see \S\ref{secradio}); the average 20~cm flux per source is $5.7\pm0.34~\mu$Jy.
In computing the 20~cm EBL, we eliminated the small number of 20~cm sources brighter than 
$300~\mu$Jy in the field, since none of these are directly detected submillimeter sources.
The 20~cm EBL is dominated by the fainter sources, and this cutoff makes only a small change 
in the results.
In Figure~\ref{frad_class}, we show the 20~cm EBL contribution from the galaxies
in the NIR sample divided into the same seven SED classes. Here, too, nearly all of 
the radio light comes from the intermediate SED classes $(2-5)$. 
These intermediate spiral galaxies have a signal of $9.1\pm0.25~\mu$Jy.

In Figure~\ref{f24_fract}, we show the fraction of sources in each SED class that 
are detected above 80~$\mu$Jy at 24~$\mu$m based on the 24~$\mu$m catalog.
Nearly all of the 24~$\mu$m sources with fluxes above $80~\mu$Jy also lie in the intermediate 
SED classes.

\begin{figure}[!h]
\epsscale{1.15}
\plotone{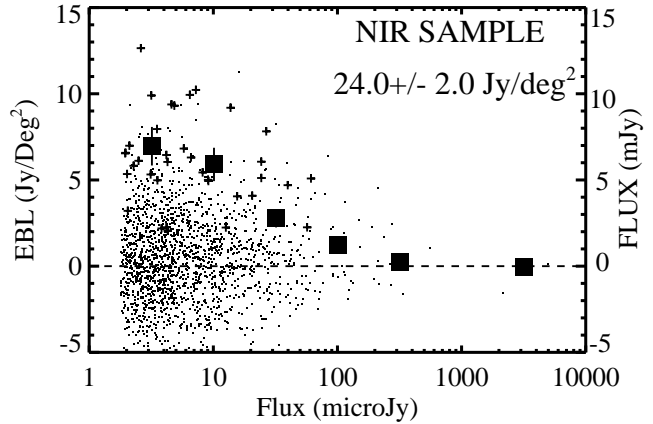}
\caption{Contributions to the 850~$\mu$m EBL from the unidentified (class 0) and
intermediate spectral type (classes $2-5$) sources (i.e., our core NIR
sample) vs. the greater of the $H$ or $3.6$~$\mu$m band fluxes. The solid squares 
show the contributions from half dex flux intervals with $1\sigma$ 
error bars. The number in the upper right corner gives the total contribution. 
The dots denote the measured fluxes of the individual sources for a y-axis in 
mJy units, and the crosses denote sources with $4\sigma$ measurements at $850~\mu$m.  
\label{base_smm}}
\end{figure}

Given that the elliptical galaxies and the very blue star-forming galaxies are only a 
source of noise, we now remove these objects and the spectroscopically-identified stars 
from the NIR sample, leaving us with 1770 sources. We refer to this sample as our core 
NIR sample. The 850~$\mu$m EBL for these sources is plotted versus the
greater of the $H$-band or 3.6~$\mu$m flux in Figure~\ref{base_smm}.
The total contribution from this final sample is 
$24.0\pm2.0$~Jy~deg$^{-2}$, or about $54-77$\% of the total 850~$\mu$m EBL.

\subsection{The Redshift Distribution of the 850~$\mu$m EBL}
\label{seczdist}

We may now use the photometric and spectroscopic redshifts to determine where in 
redshift space the 850~$\mu$m EBL arises. Nearly half of the sources (1478) in the 
full NIR sample have spectroscopic redshifts. Combining these with the photometric 
redshifts increases the identification to 3020. Only 101 sources are blended, too 
faint, or too peculiar in their SEDs to be identified.

\begin{figure}[!h]
\epsscale{1.2}
\plotone{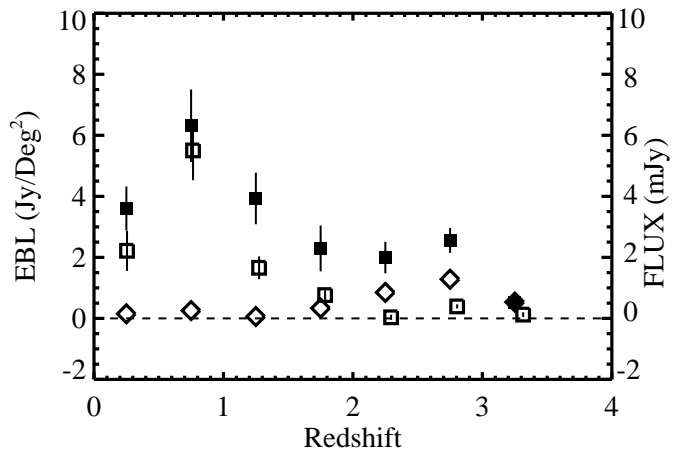}
\caption{Contributions to the 850~$\mu$m EBL from our core NIR sample with 
spectroscopic or
photometric redshifts vs. redshift.
 The solid squares 
show the contributions from each redshift interval with
$1\sigma$ error 
bars. The open squares show the contributions if we only consider sources 
with spectroscopic redshifts. The open diamonds show the
mean fluxes of the 
sources for a y-axis in mJy units.
\label{nir_smm_z}}
\end{figure}

In Figure~\ref{nir_smm_z}, we show the 850~$\mu$m EBL that arises in the core NIR 
sample (that is, the intermediate SED classes plus the unidentified sources only) 
divided by redshift interval. Here the solid squares denote the contributions 
from the sample with either photometric or spectroscopic redshifts, and the open 
squares denote the contributions from the sample with spectroscopic redshifts 
only. Open diamonds show the average flux per source with a y-axis in mJy. 
The EBL from the core NIR sample is dominated by low-redshift sources.
Indeed, $14.0\pm1.6$~Jy~deg$^{-2}$ of the EBL comes from below $z=1.5$, implying 
that about half of the 850~$\mu$m EBL is originating at these low redshifts. This 
is in striking contrast to the redshift distribution of the bright SCUBA
sources seen at higher fluxes and identified using their radio counterparts. 
It appears that at lower submillimeter fluxes, there is a substantial contribution
to the 850~$\mu$m EBL from galaxies at much lower redshifts than is the case at 
the higher submillimeter fluxes.
\vskip 0.1cm

\section{The Star-Formation History}
\label{section_evolution}

\subsection{The Star Formation Rate Density from the Core NIR Sample}

Subject to the assumed stellar mass function, the star-formation rates 
($\dot{M}$ in $M_{\sun}$~yr$^{-1}$) of sources can be estimated from their 
infrared luminosities. In our analysis, we use the formula 
$\dot{M}=1.7\times10^{-10}L_{\rm IR}/L_{\sun}$ \citep{kennicutt98}.
This is very similar to the value of
$\dot{M}=1.5\times10^{-10}L_{\rm IR}/L_{\sun}$ derived by \cite{barger00}.

The FIR luminosity can, in principle, be estimated from the $3.6-24~\mu$m
flux, and this has been done to estimate the star-formation history of
the $24~\mu$m population. However, the conversion from $3.6-24~\mu$m flux to 
FIR luminosity is complex, so this method is relatively uncertain, and 
we do not follow it here.

\begin{figure}[!h]
\epsscale{1.08}
\plotone{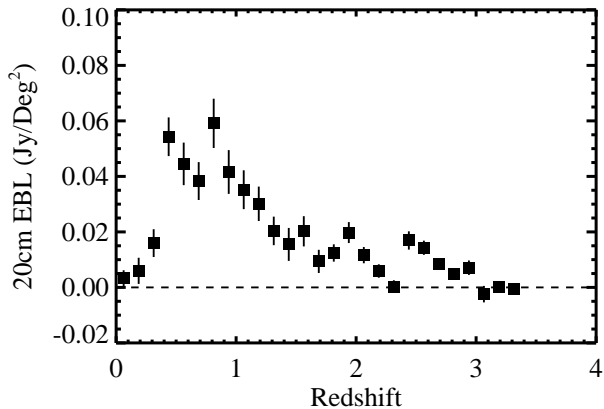}
\caption{Contributions to the 20~cm EBL from our core NIR sample with 
spectroscopic or
photometric redshifts vs. redshift. The solid squares 
show the contributions from each redshift interval with $1\sigma$ error bars. 
\label{nir_frad_z}}
\end{figure}

The direct conversion of an $850~\mu$m flux to a FIR luminosity is probably 
relatively robust for high-redshift ($z>1$) ULIRGs, and we use this in 
\S\ref{secsfrd} to estimate the star formation rate density (SFRD)
for those sources. However, as we saw in \S\ref{seczdist}, much of the
submillimeter EBL that we have been able to identify from the core NIR sample 
is at lower redshifts, and the submillimeter flux to FIR luminosity conversion
for these lower luminosity sources may have a much wider range. Thus, in order 
to compute the star-formation history of the core NIR sample, we use the 20~cm
fluxes with their robust conversion of radio power to total luminosity. 

In Figure~\ref{nir_frad_z}, we show the 20~cm EBL versus redshift. We
see a strong peak at a redshift of just below one, which then trails down 
to near zero beyond $z=2.6$. We also generated the same plot for a large 
number of samples of randomized positions in the field containing the same 
number of sources. The random realizations average to zero, are fully 
consistent with the range expected from the statistical noise, and show 
no systematic effects. In addition, we computed the 20~cm EBL with
redshift for only the sources in the \citet{richards00} catalog with its 
limiting flux of 40~$\mu$Jy. As might be expected, since this sample is 
considerably brighter in flux than the stacking sample, we found a lower
peak redshift, with a value of $z\sim 0.5$.  

We note that 
Figures~\ref{nir_smm_z} and \ref{nir_frad_z} imply an unusually large 
submillimeter-to-radio flux ratio at $z\sim1$.  This is consistent with 
cool dust in local low-luminosity galaxies \citep[e.g.,][]{vlahakis05} or with
a great amount of high-redshift background light being lensed by low-redshift 
sources \citep[e.g.,][]{almaini05}. 

We next translated the 20~cm EBL into a bolometric luminosity density for each 
redshift interval. We used the center redshift of the interval to compute the 
radio power per unit area, which
we then converted to a total luminosity using the FIR-radio correlation (e.g.,
\citealt{condon92}). We next computed the volume in each redshift bin for the 
unit area, thereby forming the total luminosity density per unit volume at that 
redshift.

\begin{figure}[!ht]
\epsscale{1.15}
\vskip 0.5cm 
\plotone{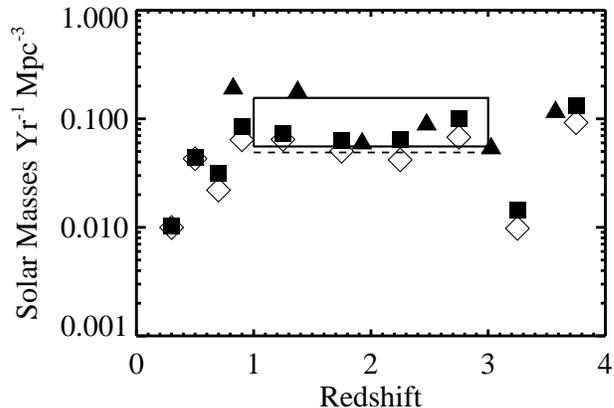}
\caption{Submillimeter SFRD vs. redshift. The solid squares show the 
SFRD derived from the 20~cm EBL contributions of our core NIR sample.
(The $1~\sigma$ errors are smaller than the symbol sizes.)
The open diamonds show the same results when sources with a $2-8$~keV
luminosity above $10^{42}$~ergs~s$^{-1}$ (sources containing AGNs)
are excluded. The solid triangles show the SFRD computed using the 
850~$\mu$m EBL. The dashed horizontal line shows the SFRD computed 
assuming all of the sources with 850~$\mu$m fluxes above 4~mJy  
lie at $z=1-3$, as suggested by the bright source identifications.
The rectangular region denotes the SFRD from the 
remaining submillimeter EBL that is not accounted for by our NIR sample, 
assuming that it also lies in the redshift interval $z=1-3$. 
The range corresponds to the uncertainty in the 850~$\mu$m EBL. The 
maximum total submillimeter SFRD in this redshift range is then the 
sum of the rectangle and the measured points.
\vskip 0.1cm
\label{mdot}}
\end{figure}

In Figure~\ref{mdot}, we show the SFRD of the core NIR sample obtained by 
converting the luminosity density to a SFRD using the \citet{kennicutt98} relation. 
The SFRD \emph{(solid squares)} shows a rapid rise to $z=0.8$, and then it 
flattens at higher redshifts. Since some part of the light may arise from AGNs 
rather than star formation, we also show in Figure~\ref{mdot}
the SFRD from only the sources that have X-ray luminosities less than 
$10^{42}$~ergs~s$^{-1}$ \emph{(open diamonds)}. 
This has only a small effect, so unless the correction for X-ray obscured AGNs 
is much larger, we can ignore the AGN contamination at our current level
of accuracy. 

At each redshift, the measured SFRD is a lower bound on the total submillimeter
SFRD, since the core NIR sample is not a complete mapping of the submillimeter 
star-formation history. In particular, it may be biased to lower-redshift sources
and preferentially miss the higher-redshift star formers. In \S\ref{secsfrd}, we 
assess the maximum corrections that are possible from the residual $850~\mu$m 
light that the core NIR sample omits.

\subsection{Limits on the Star Formation Rate Density at $z=1-3$}
\label{secsfrd}

For the higher-redshift sources, where the galaxies are near-ULIRGs,
we may make a direct estimate of the SFRD from the submillimeter light.
We again assume that the luminosities are dominated by star formation. The 
infrared luminosities of the submillimeter sources can be estimated from their  
850~$\mu$m fluxes, redshifts, a plausible dust temperature ($T_d$), and a dust 
emissivity index ($\beta$) in the submillimeter. To make this conversion,
we adopted the dust model in \citet{yun02} and their values of 
$T_d=58\pm9$ and $\beta=1.32\pm0.17$ based on luminous and 
ultraluminous starbursts. This conversion is very similar to the one
obtained for $T_d=47$ and $\beta=1.0$ (Arp~220), which has often
been used in the past \citep[e.g.,][]{barger00}.

\begin{figure*}[!t]
\epsscale{0.7}
\vskip 0.2cm
\plotone{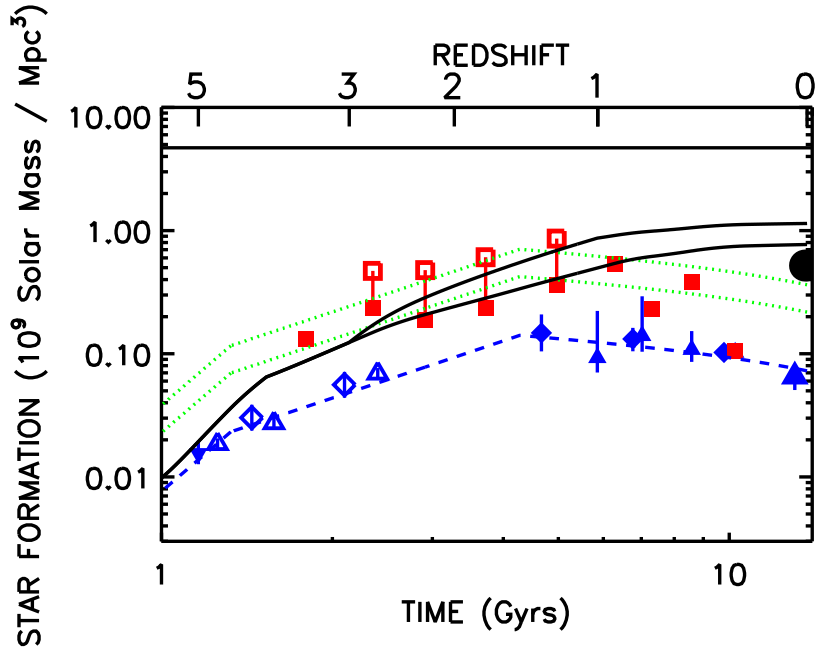}
\caption{Comparison of the star formation history vs. cosmic time 
from our FIR determinations with that from UV samples. We plot 
the star formation rate density computed for a Salpeter IMF extending to 
$0.1~{\rm M}_\sun$ multiplied by the cosmic time, since this shows 
more clearly what fraction of the baryonic mass in stars is created at 
any time. The triangles and diamonds show the star formation that
is directly seen at  rest-frame UV wavelengths. At late times, we show
the local {\em Galaxy Evolution Explorer} (GALEX) determination of 
Wyder et al.\ (2005) {\em (large solid triangle)}, the GALEX
determinations of Schiminovich et al.\ (2005) {\em (small solid triangles)}, 
and the ground-based determinations of Wilson et al.\ (2002) 
{\em (small solid diamonds)}, which are the most accurate
measurements near $z=1$. At intermediate times, we show the 
determinations of Steidel et al.\ (1999) {\em (open diamonds)},
Bouwens, Broadhurst, \& Illingworth (2003) {\em (small open triangles)},
and Iwata et al.\ (2003) {\em (solid upside-down triangle)}.
The dashed curves show a parameterized fit.  The dotted green curves 
show the total star formation that would be inferred if we were to apply reasonable 
extinction corrections to the UV light (upper curve = 5, lower curve =3). 
The solid red squares show this paper's FIR determinations, and the open 
red squares show our maximal corrections for incompleteness in the $z=1-3$ 
range. There is broad agreement of the FIR determinations 
with the dust-corrected UV determinations, 
though the populations giving rise to the FIR light are somewhat disjoint
from those giving rise to the UV light. The horizontal line shows the cosmic 
baryon density. The solid curves show the cumulative star formation. 
The solid black circle shows the present-day stellar baryon density
estimated by Cole et al.\ (2001). \vskip 0.2cm}
\label{sfh}
\end{figure*}

We first computed the SFRD from the 850~$\mu$m EBL determined
as a function of redshift for the core NIR sample. We computed the
FIR luminosity per unit area in each redshift bin using the adopted dust model. 
Next, we computed the volume corresponding to the redshift interval. Finally,
we converted the FIR luminosity per unit volume to a SFRD 
using the \citet{kennicutt98} relation. The result is plotted in Figure~\ref{mdot} 
as the solid triangles. At $z>1$, these results agree strikingly well with the 
20~cm-determined SFRD shown by the solid squares. This is not a foregone
conclusion, since we are using two independent data sets for the computation 
and two different (though hopefully consistent) calibrations for the 
conversion to luminosity. (The FIR-radio correlation in the case of the
20~cm conversion, and the \citet{yun02} dust model for the submillimeter
conversion.) At $z<1$, the 850~$\mu$m-determined SFRD is higher 
than the 20~cm determined SFRD, which is expected, since the adopted luminosity 
conversion may no longer be appropriate if the sources are no longer
near-ULIRGs.

We next computed the SFRD assuming that all of the EBL from sources with 850~$\mu$m fluxes 
greater than 4~mJy is in the $z=1-3$ range, where the identified sources of 
Table~\ref{tab1} lie.
The fact that deep radio imaging only 
detects $\sim60$\% of the bright SCUBA sources suggests that the remaining 
40\% may be at redshifts greater than $z\sim3$; thus, we may be overestimating 
this contribution. 
This result is plotted in Fig.~\ref{mdot} as the dashed line.
The SFRD measured in this way is not independent of the SFRD measured in the core 
NIR sample, since the bright 850~$\mu$m sources may already be contained in the 
core NIR sample. Indeed, the directly determined value lies below the core NIR 
SFRD at these redshifts. This is because, while we are making the extreme assumption 
that all of the bright sources lie in this redshift range, only a small fraction of 
the 850~$\mu$m EBL lies at these bright fluxes (about 7~Jy~deg$^{-2}$). 

Finally, the maximum completeness correction for the submillimeter SFRD in 
this redshift interval can be determined from the residual 850~$\mu$m 
EBL that is not identified by the core NIR sample, assuming that it
all lies at $z\sim1-3$. Most likely only a part of this missing
light will come from this redshift range and at least some
may come from higher redshift sources where the counterparts
at other wavelengths will be much fainter.
This remaining light is approximately $7-21$~Jy~deg$^{-2}$ 
in 850~$\mu$m EBL, depending on the total EBL value used. 
The SFRD from the residual submillimeter EBL is shown by the 
rectangular area in Figure~\ref{mdot} and can be added
to the SFRD from the core NIR sample to obtain the maximum
possible submillimeter SFRD at each redshift. 

The SFRD derived from the core NIR sample using the radio
data is $0.09~M_{\sun}$~yr$^{-1}$~Mpc$^{-3}$
at $z=1$ and is almost identical to that at $z=2-3$. Even the maximum 
completeness correction of $0.05~M_{\sun}$~yr$^{-1}$~Mpc$^{-3}$ to
$0.15~M_{\sun}$~yr$^{-1}$~Mpc$^{-3}$ can only result in a star-formation 
history that is rising slowly above $z\sim 0.8$. When integrated through 
time, it appears that the majority of the star formation occurs around 
$z=1$. Thus, previous claims that the submillimeter SFRD is strongly 
peaked at $z=2-3$ appear to be incorrect.

\subsection{Comparison with the UV-Determined Star Formation Rate Density}
\label{secuv}

In Figure~\ref{sfh}, we compare the SFRD determined here with the SFRD determined 
from UV observations. Both are computed consistently with a Salpeter IMF extending 
to $0.1~{\rm M}_{\sun}$. However, rather than plotting $\dot{\rho}$ versus 
redshift, as is usually done, we plot the quantity $\dot{\rho} \times t$ versus the cosmic 
time, $t$. The advantage of this display is that we can see more directly how many stars 
are formed at a given time. We use diamonds and triangles to show the direct star formation 
determinations from the UV light without any correction for extinction. We use
solid squares to show our directly measured FIR star formation, and we use open squares
to show our maximal corrections for the missing EBL, if it is formed in the $z=1-3$ redshift 
range.

The FIR star formation is about a factor of 3 to 5 higher than the uncorrected UV star formation,
depending on the correction applied for the missing $850~\mu$m light {\em (dotted
curves)}. The lower value corresponds to the directly measured light, and the upper value 
to the maximally corrected light. This is consistent with the usual dust corrections 
applied to the UV star formation rates, but it should be noted that the samples are 
somewhat disjoint, in that the blue star-forming galaxies (our template classes 6 and 7)
contribute substantially to the UV star formation but not to the FIR star formation. 
Thus, the extinction corrections must be higher in the galaxies 
that produce the bulk of the FIR light and lower in the blue star-forming galaxies.

We have also computed the total amount of stars formed as a function of time by
combining the FIR and UV star-formation histories and integrating
with respect to cosmic time.  Above $z=4$, where we do not have any information
about the FIR light, we have assumed that the total star
formation is five times the UV star formation, but the results at later times are quite insensitive 
to this assumption, since only a very small part of the total star formation occurs at these early 
times. The cumulative total is shown by the solid curves, where the lower curve 
corresponds to the directly measured FIR star formation and the upper curve to the 
maximally corrected FIR star formation. At $z=0$, the curves match suprisingly well 
to the local determination of the stellar mass density \citep{cole01}, which is shown by 
the large solid circle in Figure~\ref{sfh}. The agreement is best for the lower estimates,
but even for the maximally corrected case, the curve is only slightly high.
For the directly measured case, half of the star formation occurs
at $z<1.3$, while in the maximally corrected case, this rises to $z<1.45$.

\section{Summary}
\label{secsummary}

We have obtained accurate redshifts for the sources in the GOODS-N area
using existing spectroscopic redshifts and improved photometric redshifts 
from NIR and MIR data. The radio-identified bright ($>2$~mJy) SCUBA
sources in this area are in the redshift range $z\sim1-3$ and have a median 
redshift of $z=2.5$, consistent with previous radio and spectroscopic surveys.

However, we used a stacking analysis to show that much of the 850~$\mu$m EBL 
is in fact traced by a NIR sample constructed from sources with fluxes greater 
than $1.8~\mu$Jy in either the $H$ or $3.6~\mu$m bands. We showed that 
much of this light arises from galaxies with intermediate spectral types
at $z<1.5$. Thus, many of the fainter submillimeter sources that give rise 
to most of the 850~$\mu$m EBL are at lower redshifts and lower luminosities 
than the bright submillimeter sources that are detected directly.

Finally, we used a stacking analysis to estimate the average 20~cm EBL produced
by the unidentified or intermediate spectral type galaxies in our NIR sample 
as a function of redshift, from which we determined the SFRD.
We found that this SFRD evolves rapidly between $z=0$ and $z=0.8$, after
which it becomes approximately flat. Using the submillimeter data directly, we 
then calculated a submillimeter based SFRD at $z\sim 1-3$  which agrees closely
with the radio based SFRD.
In addition, by assuming that all of the submillimeter EBL that is not accounted 
for by our NIR sample is also at these redshifts, we put an upper bound on the 
SFRD at $z\sim 1-3$. Even with this maximum completeness correction, we found
consistency with a nearly flat or slowly rising extrapolation of the SFRD from $z\sim1$. 
We conclude that the majority of the star formation traced by the submillimeter 
light comes from redshifts near one rather than at the higher redshifts 
that have been favored until now.

\vskip -0.3cm

\acknowledgments
We thank P.~Capak for useful discussions about photometric redshifts and
the referee, S. Serjeant, for helpful suggestions on improving the paper.
We gratefully acknowledge support from NSF grants AST 04-07374 (L.L.C.)
and AST 02-39425 (A.J.B.), the University of Wisconsin Research Committee
with funds granted by the Wisconsin Alumni Research Foundation,
the Alfred P. Sloan Foundation, and the David and Lucile Packard
Foundation (A.J.B.).

\end{document}